\documentclass[aps,prl,twocolumn,showpacs]{revtex4}

\usepackage{amsmath}
\usepackage{graphicx}
\usepackage{mathdots}

\usepackage{epsf}

\newcommand \bea {\begin{eqnarray} }
\newcommand \eea {\end{eqnarray}}

\newcommand{\beg}{\begin{equation}}
\newcommand{\en}{\end{equation}}

\newcommand{\eps}{\varepsilon}

\newcommand{\re}[1]{(\ref{#1})}

\begin{document}

\newcommand{\dg}{^{\dagger }}
\newcommand{\vk}{\mathbf k}
\newcommand{\vq}{\mathbf q}
\newcommand{\vp}{\mathbf p}
\newcommand{\vl}{\mathbf l}
\newcommand{\bfr}{\mathbf r}
\newcommand{\sinp}{\sin\left(\frac{\theta_\vp}{2}\right)}
\newcommand{\veps}{\varepsilon}
\newcommand{\up}{\uparrow}
\newcommand{\dn}{\downarrow}

\title{Cooper pair turbulence in atomic Fermi gases}

\author{M. Dzero$^{1,2}$, E. A. Yuzbashyan$^2$, B. L. Altshuler$^1$}
\affiliation{$^1$ Department of Physics, Columbia University, New York, NY 10027, USA\\
$^2$Center for Materials Theory, Rutgers University, Piscataway, NJ 08854, USA}

\begin{abstract}
 We investigate the stability of spatially uniform solutions for the collisionless dynamics of a fermionic superfluid. We demonstrate that, if the system size is larger than the superfluid coherence length, the solution characterized by a periodic in time order parameter is unstable with respect to spatial fluctuations. The instability is due to the parametric excitations of pairing modes with opposite momenta. The growth of spatial modulations is  suppressed by nonlinear effects resulting in a state characterized by a random superposition of wave packets of the superfluid order parameter.
We suggest that this state can be probed by spectroscopic noise measurements. 
\end{abstract}

\pacs{05.30.Fk, 32.80.-t, 74.25.Gz}

\maketitle


How does a homogeneous interacting many-body system develop spatial
modulations as a result of a uniform quench? In general, this can happen  due
to an interplay between  the extra energy introduced by the quench and an intrinsic coupling of
the degrees of freedom at various length scales \cite{Ruutu1995,Warner2005,oil2006,BEC2006}.
By the very nature of the problem, the energy distribution of these degrees of freedom is initially
far from a state of thermodynamic equilibrium.
This situation is common  in a nonlinear medium and can be described
by the term "wave turbulence" \cite{Lvov1975}.  In the case when one of the parameters of the medium or an applied field is periodically varied in time, wave turbulence due to a parametric instability can develop. Examples of such a phenomenon include the decay of high-frequency electric field into
Lagmuir and ion-sound waves in plasma \cite{Ruben1973} and the spin-wave instability in an
rf-magnetic field  in dielectric ferromagnets \cite{Lvov1976}. A natural question is then whether a quench can lead to a parametric excitation of spatial modes.


Oscillating homogeneous fields can be generated by a uniform quench without an
external drive. One of the recent examples where this takes place is a fermionic superfluid    quenched
 by a sudden change   of the pairing strength  \cite{Levitov2004,classify}.  There are two types of asymptotic states the system can reach   depending on the strength of the initial perturbation  \cite{Levitov2006,Emil2006}.  The first type has a constant value $\Delta(t)=\Delta_s$, while the second one is characterized by periodic $\Delta(t)$.    This analysis does not take into account the pair breaking processes and is thus valid only for $t<\tau_\varepsilon$, where $\tau_\varepsilon$ 
 is the quasiparticle relaxation time.
Moreover, the emerging asymptotic states are spatially uniform:
order parameter evolution was obtained as a solution of an effectively zero
dimensional problem. Results of Refs.~\cite{Levitov2006,Emil2006}
are thus valid   when the system size is smaller than the
coherence length, $L<\xi$. The description of the order parameter evolution for
$L>\xi$  requires an additional investigation.

\begin{figure}[h]
\includegraphics[width=2.8in]{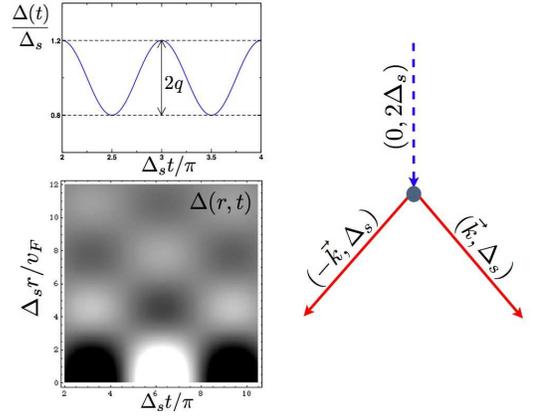}
\caption{Asymptotic state with the periodic order parameter $\Delta(t)=\Delta_s[1+q\cos(2\Delta_st)]$ 
(top left) is unstable with respect to the parametric excitation of two modes with opposite momenta (right panel).
Frequency of each mode is half the oscillation frequency of
the homogeneous order parameter. Initial exponential growth of the parametric instability is followed by a transient regime after which the condensate reaches a spatially inhomogeneous post-threshold state
(bottom left). The value of the order parameter
increases from white to black.}
\label{Fig1}
\end{figure}

In this paper, we perform a stability analysis of
the solutions for the wave function and order parameter
obtained in Refs. \cite{Levitov2006,Emil2006}
with respect to spatial fluctuations. We find that
the asymptotic states with constant order parameter remain stable, while the state
with periodic $\Delta(t)$ does not. The physical origin
of the instability lies in the possibility of parametric excitation of the spatially
modulated pairing modes, Fig. \ref{Fig1}. In a homogeneous medium,
the periodic (in time) order parameter can be considered as an energy pump
allowing a coupling between two spatial   modes with opposite momenta,
and, at the same time, providing enough energy to overcome the damping due to the scattering
of Cooper pairs. Subsequent scattering effects of the resonant modes limit the initial
exponential growth and result in a state with a \emph{spatially inhomogeneous}
order parameter. We demonstrate that as the process of the parametric instability develops, 
the energy of the  homogeneous state is transferred into that of the pairing wave packets
with the typical size of the order of the coherence length, $\xi$, signaling the onset of the Cooper pair turbulence. Since the amplitudes of the wave packets are essentially random,
we suggest that this state can be experimentally probed by the noise measurements.

Consider the   state with a periodically varying  order parameter.
Analytically, $\Delta(t)$ is described by the Jacobi elliptic
function dn \cite{Levitov2004}. Here we assume that the amplitude of the  oscillations is small. This allows us to keep only the first two terms in the Fourier series for dn:
\beg\label{dn}
\Delta(t)=\Delta_s[1+q\cos(2\Delta_st)], \quad q\ll 1.
\en
We note that the non-dissipative  dynamics within the BCS model is described
by the Bogoliubov-de Gennes  equations, which can be cast into the form
of equations of motion for classical vector variables ${\vec s}_\vp$:
$\dot{\vec s}_\vp={\vec b}_\vp\times{\vec s}_\vp$. Here ${\vec b}_\vp=2(-\Delta(t),0,\varepsilon_\vp)$ plays the role of an external magnetic field. The periodic external field makes it possible for
the parametric instabilities to develop.
Given that the asymptotic state \re{dn} is robust against homogeneous
perturbations  \cite{Levitov2004,classify},
it seems natural to look for an instability with respect to spatial fluctuations.
In particular, we will investigate the conditions of the parametric
decay of the homogeneous pairing mode \re{dn} with an energy $2\Delta_s$ into two pairing modes with energies and momenta $(\omega_1,\vk_1)$
and  $(\omega_2,\vk_2)$, Fig. 1. In a continuous medium the energy and
momentum have to be conserved:
$\vk_1+\vk_2=0$ and $2\Delta_s=\omega_1+\omega_2$.
These type of instabilities appear in various physical systems,
see e.g. Ref.~\cite{Lvov1975} for an extensive review.

To analyze the stability of the asymptotic state \re{dn}, we
employ the Bogoliubov-de Gennes (BdG) equations\cite{clean}:
\beg\label{BdG}
\begin{split}
&i\dot{u}_{\vp}(\bfr,t)=\hat{\xi}u_{\vp}(\bfr,t)+
\Delta(\bfr,t)v_{\vp}(\bfr,t), \\
&i\dot{v}_{\vp}(\bfr,t)=-\hat{\xi}v_{\vp}(\bfr,t)+
{\bar \Delta}(\bfr,t)u_{\vp}(\bfr,t).
\end{split}
\en
Here $\hat{\xi}=-{\vec \nabla}^2/{2m}-\mu$, $\mu$ is the chemical potential and the order parameter
$\Delta(\bfr,t)$ is
\beg
\label{self}
\Delta(\bfr,t)=g\sum\limits_{\vp}u_{\vp}(\bfr,t){\bar v}_{\vp}(\bfr,t),
\en
where $g$ is the BCS coupling constant. We linearize Eqs. (\ref{BdG}) with respect to
the deviations $\phi_\vp(\bfr,t)$ and $\psi_\vp(\bfr,t)$ from the homogeneous solution
$U_\vp(t)$ and $V_\vp(t)$:
\beg\label{deviate}
\left[
\begin{matrix}
u_{\vp}(\bfr,t) \\
v_{\vp}(\bfr,t)
\end{matrix}
\right]=\left[
\begin{matrix}
U_{\vp}(t)+\phi_\vp(\bfr,t) \\
V_{\vp}(t)+\psi_\vp(\bfr,t)
\end{matrix}
\right]e^{i\vp\cdot\bfr}.
\en
The order parameter (\ref{self}) also contains a small inhomogeneous part
$\Delta(\bfr,t)=\Delta(t)+\delta\Delta(\bfr,t)$. Plugging (\ref{deviate}) into (\ref{BdG}) and  Fourier transforming with respect to the spatial coordinates, we find
\beg\label{linearized}
\begin{split}
i\partial_t\left[
\begin{matrix}
{\phi}_\vp(\vk,t) \\
{\psi}_\vp(\vk,t)
\end{matrix}
\right]=&\left(
\begin{matrix}
\eps_{\vp+\vk} ~~\Delta(t) \\
\Delta(t) ~~ -\eps_{\vp+\vk}
\end{matrix}\right)\left[
\begin{matrix}
{\phi}_\vp(\vk,t) \\
{\psi}_\vp(\vk,t)
\end{matrix}
\right]\\ &+
\left(\begin{matrix}
0 ~~\delta\Delta(\vk,t) \\
\delta{\bar \Delta}(-\vk,t) ~~ 0
\end{matrix}\right)
\left[
\begin{matrix}
{U}_\vp(t) \\
{V}_\vp(t)
\end{matrix}
\right]
\end{split}
\en
The time dependance of $U_\vp(t)$ and $V_\vp(t)$ (see Ref.~\cite{Dzero2007})  suggests
that for linear corrections (\ref{deviate}) we write  $\phi_\vp(\vk,t)=
a_\vp(\vk,t)e^{i\xi_\vp t}+b_{\vp}(\vk,t)e^{-i\xi_\vp t}$ and
$\psi_\vp(\vk,t)=\tilde{a}_\vp(\vk,t)e^{i\xi_\vp t}+\tilde{b}_{\vp}(\vk,t)e^{-i\xi_\vp t}$ with $\xi_\vp=\sqrt{\varepsilon_\vp^2+\Delta_s^2}+O(q)$ and $\varepsilon_\vp=\frac{\vp^2}{2m}-\mu$.
We  also write
\beg\label{Delta1k}
\delta\Delta(\vk,t)=C_\vk(t) e^{i\Delta_st}+\widetilde{C}_\vk(t) e^{-i\Delta_st},
\en
where $C_\vk(t)= c_\vk e^{\nu(t-t_0)}$, $\widetilde{C}_\vk(t)=\tilde{c}_\vk e^{\nu(t-t_0)}$, 
$\nu$ determines the growth rate, and $t_0$ is a
time scale when (\ref{dn}) is reached
(in what follows we set $t_0=0$). In the expression (\ref{Delta1k})
we have neglected the higher harmonics $e^{\pm i3\Delta_st}, e^{\pm i5\Delta_st}$ etc.
This  is justified for $q\ll 1$, since their inclusion  yields
higher order in $q$ corrections to the growth rate $\nu$ and to the order
parameter amplitudes \cite{LandafshitzI}. In the linear corrections to the Bogoliubov amplitudes
(see above) we also keep only the lowest harmonics  $\omega=\pm\Delta_s$,
i.e. $a_\vp(\vk,t)\to (a_{1,\vp}(\vk)e^{i\Delta_st}+
a_{-1,\vp}(\vk)e^{-i\Delta_st})e^{\nu t}$, $b_\vp(\vk,t)\to (b_{1,\vp}(\vk)e^{i\Delta_st}+
b_{-1,\vp}(\vk)e^{-i\Delta_st})e^{\nu t}$ etc.

Next, we  express the Bogoliubov
amplitudes in terms of $c_\vk$ and $\tilde{c}_\vk$
by equating the coefficients in front of $e^{\pm i\Delta_st}$ in Eq.~(\ref{linearized}).
The resulting amplitudes are substituted into the self-consistency equation (\ref{self}).
One obtains a linear system for the variables $c_\vk, \tilde{c}_{-\vk}^*, c_{-\vk}^*$ and $\tilde{c}_{\vk}$.
Expressing $\tilde{c}_{\vk}, \tilde{c}_{-\vk}^*$ in terms of $c_\vk$ and $c_{-\vk}^*$,
we  derive
\beg\label{stabilityeqs}
(\omega_\vk+i\gamma_\vk)c_\vk+h_\vk c_{-\vk}^*=0,
~~(\omega_\vk-i\gamma_\vk)c_{-\vk}^*+h_\vk c_{\vk}=0,
\en
where $\omega_\vk$, $\gamma_\vk$, and $h_\vk$ are nonlinear functions of the growth rate $\nu$.
Eqs.~(\ref{stabilityeqs}) can be  interpreted
as the equations of motion for a classical field $c_\vk$ \cite{Lvov1975}. 
Then, $\omega_\vk$ has the meaning of
the excitation spectrum of this field and $h_\vk\sim O(q)$
stands for the pumping amplitude, which gives rise to a parametric instability.
Finally, $\gamma_\vk$ describes the  damping of the parametric modes due
to the intrinsic relaxation processes.

Nonzero solutions of  Eqs.~(\ref{stabilityeqs})  for $\nu(\vk)$ exist provided
$\omega_\vk^2(\nu)=h_\vk^2(\nu)-\gamma_\vk^2(\nu)$.
Thus, the stability analysis is reduced to the solution of the nonlinear equation for $\nu(\vk)$.
We have analyzed this equation numerically and present the results on Fig. 2.
We find that the instability region is centered around $k_m\approx 1.6k_\xi$ and has a width
$\delta k\approx 1.2\sqrt{q}k_\xi$, where $k_\xi=\Delta_s/v_F=1/\xi$ is the coherence wave vector.
From (\ref{stabilityeqs}) it follows that for a fixed $q$ the parametric growth will be suppressed as soon as the  energy  pumped into the system fully goes into dissipation. This condition determines the maximum growth rate $\nu_m$, i.e.
$\gamma_{\vk}(\nu_m)=h_{\vk}(\nu_m)$. Our estimate yields $\nu_m\approx 2q\Delta_s$. Lastly, we have also verified that asymptotic states with constant order parameter remain stable with respect to the spatial fluctuations of the above type.
\begin{figure}[h]
\includegraphics[width=2.8in]{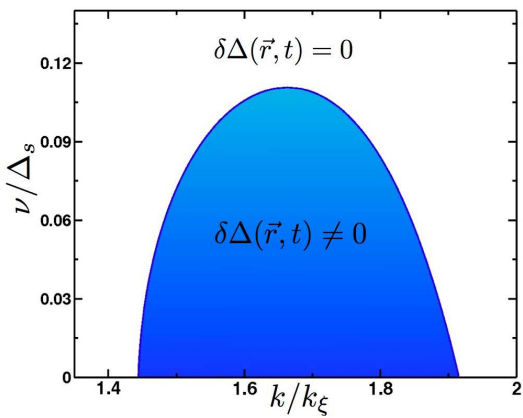}
\caption{Region of the parametric instability of the homogeneous $\Delta(t)$ (\ref{dn})
with respect to generation of the pairing modes with opposite momenta ($\vk,-\vk$).
Instability growth rate
is plotted for $q=0.05$ in the units of $\Delta_s$ (see Eq. (\ref{dn})) and momentum is
in the units of $k_\xi=\Delta_s/v_F$. The maximum rate is reached at $\nu_m\approx2q\Delta_s$. For small $q$ the shape of the instability curve
is $\nu(k)\approx \nu_m-2\Delta_s(k-1.6k_\xi)^2/k_\xi^2$.}
\label{Fig2}
\end{figure}

The initial growth of the parametric instability (\ref{initgrowth})
will be limited by nonlinear effects which lead
to the transient behavior with subsequent transition into a post-threshold state.
The latter is defined as a state in which Fourier components of the order parameter
(\ref{Delta1k}) are time independent, $C_{\vk}(t)=c_\vk$ and $\widetilde{C}_\vk(t)=\tilde{c}_\vk$. 
Below we  focus on finding the resulting
post-threshold state of the condensate. From the linear analysis we have seen that the
fastest growing modes are the ones with
a certain magnitude of the momentum. Thus, in the BdG equations (\ref{BdG})
among the nonlinear in powers of $c_\vk, \tilde{c}_\vk$ terms we  keep the resonant ones
with frequencies $\omega=\pm\Delta_s$ and momenta $|\vk|=|\vk'|=k_{s}$,
where $k_s$ is a new post-threshold state momentum to be determined below.
The resulting set of nonlinear in $c_\vk$ equations for the order parameter
amplitudes can be written as Eq. (\ref{stabilityeqs}) with renormalized coefficients
\beg\label{deltack}
\begin{split}
&\omega_\vk\to\Omega_\vk=
\omega_\vk(0)+\sum\limits_{|\vk'|=k_s}T_{\vk\vk'}|c_{\vk'}|^2, \\
&h_\vk\to P_\vk=h_\vk(0)+\sum\limits_{|\vk'|=k_s}S_{\vk\vk'}c_{\vk'}c_{-\vk'},
\end{split}
\en
where $T_{\vk\vk'}$ and ${S}_{\vk\vk'}$ are the scattering matrix elements. They
vary slowly on a scale of $k_\xi$ and are
almost independent of the angle between $\vk$ and $\vk'$.
In what follows we neglect the $k$-dependence in the scattering matrix 
elements, $T_{\vk\vk'}=T$ and $S_{\vk\vk'}=S$.
Note that each contribution in (\ref{deltack}) is either phase independent or depends on a sum
of the two phases of $c_\vk$ and $c_{-\vk}^{*}$.
This can be interpreted as follows. There are two physical processes which
limit the parametric excitations: one has to do with the reduction of the absolute value of the
amplitudes, while the other is related to the phase decoherence of the two pairing modes with
opposite momenta. In its spirit approximation (\ref{deltack}) is similar to the mean-field BCS model where only diagonal in momentum terms are kept in the interaction.
The inclusion of off-diagonal terms in Eq. (\ref{deltack}) is expected to cause a broadening of the
post-threshold state momentum $\delta k\sim\sqrt{q}k_\xi$, see Fig. 2.

To determine the parameters of our post-threshold state, we insert $c_\vk=|c_\vk|e^{i\alpha_\vk}$,
and Eqs. (\ref{deltack}) for $\Omega_\vk$ and $P_\vk$ into (\ref{stabilityeqs}).
The post-threshold state momentum $k_s$ is determined by the condition that the magnitude of
the pumping field $|P_\vk|$ does not exceed the damping $\gamma_\vk$ for any $k$.
As a result we have $\Omega_{\vk_s}=0$. Phase $\Psi_s=\alpha_{\vk}+\alpha_{-\vk}$
and amplitude $|c_{k_s}|$ are given by
$\sin\Psi_s=\gamma_{k_s}/h_{k_s}$ and $|c_{k_s}|^2=h_{k_s}\cos\Psi_{s}/|S|$
(the corresponding expressions for $|\tilde{c}_{k_s}|$ and
$\widetilde{\Psi}_{s}=\tilde{\alpha}_{\vk}+\tilde{\alpha}_{-{\vk}}$ can be derived
similarly). We obtain
\beg\label{steadyOP}
\begin{split}
\Delta({\vec r},t)=&\Delta_s+{\sqrt{q}\Delta_s}{c_s}\sum\limits_{|\vk|=k_s}
e^{i\vk\cdot\bfr}\\&\times\left[e^{i(\alpha_{\vk}+\Delta_st)}+
w_se^{i(\tilde{\alpha}_{\vk}-\Delta_st)}\right],
\end{split}
\en
where $k_s\approx1.73k_\xi$, $c_{s}\approx 0.77$ and $w_{s}\approx 0.95$ for $\Delta_s=0.1\mu$.
Note that the post-threshold momentum $k_s>k_m\approx 1.6k_\xi$, 
i.e. the energy cascades to smaller length scales, as expected of turbulent behavior. 

The individual phases $\alpha_{\vk}$ and  $\tilde{\alpha}_{\vk}$
cannot be determined within the diagonal approximation (\ref{deltack}).
In a continuous medium, one can treat them as random variables.
For the correlators we take $\langle e^{i\alpha_{\vk}}\rangle=0$, $\langle e^{i\alpha_{\vk_1}}e^{i\tilde{\alpha}_{\vk_2}}\rangle=0$ and $\langle e^{i\alpha_{{\vk}_1}}e^{i\alpha_{{\vk}_2}}\rangle=
\delta_{{\vk}_1,-{\vk}_2}^{(3)}e^{i\Psi_s}$, where $\langle ... \rangle$ stands for
averaging over the phase distribution.

To get further insight into the nature of the post-threshold state,
consider the following choice for the phases $\alpha_\vk=\Psi_s/2-\vk\cdot\bfr_0$ and
$\tilde{\alpha}_\vk=\widetilde{\Psi}_s/2-{\vk}\cdot\bfr_0$.
It leads to a spherically symmetric wave packet, a "bubble",
with a periodic amplitude $A(t)$
\beg\label{steadyOP1}
\begin{split}
&\Delta({\vec r},t)=\Delta_s+
\frac{\sqrt{q}\Delta_sc_s\sin(k_s|{\vec r}-{\vec r}_0|)}{k_s|{\vec r}-{\vec r}_0|}A(t), \\
&A(t)=e^{i(\frac{\Psi_s}{2}+\Delta_st)}+
w_se^{i(\frac{\widetilde{\Psi}_s}{2}-\Delta_st)}.
\end{split}
\en
In general we obtain a linear combination of these bubbles (\ref{steadyOP1})
by writing  $e^{i\alpha_\vk}=\int f(\bfr_0)
e^{-i\vk\cdot\bfr_0}d^3\bfr_0$ and similarly for $e^{i\tilde{\alpha}_\vk}$.
Then, Eq. (\ref{steadyOP}) can be viewed as a superposition of wave packets of the
form (\ref{steadyOP1}) centered at different ${\vec r}_0$ with random
amplitudes $A({\vec r}_0,t)$. This suggests that the parametric instability results in a
random distribution of the wave packets.

It is  instructive to compare (\ref{steadyOP1}) with $\delta\Delta({\vec r},t)$ at the linear
stage of the parametric instability, which can be derived using our result for
$\nu(k)$ (see Fig. 2) and Eqs.~(\ref{Delta1k},\ref{stabilityeqs}). Taking $c_\vk=Ce^{-i(\vk\cdot\bfr_0+\Delta_s\tau_\vk)}$ and  $\tilde{c}_\vk=c_{-\vk}^*$, we obtain
\beg\label{initgrowth}
\delta\Delta({\vec r},t)\approx
\frac{Ce^{\nu_m t}\cos[\Delta_s(t-\tau)]}{\sqrt{\Delta_st}}\frac{\sin(k_m R)e^{-R^2/l^2(t)}}{k_mR},
\en
where $l(t)\approx\xi\sqrt{\Delta_st}$,  $R=|{\vec r}-{\vec r}_0|$,  $C$ is a constant, and $e^{i\Delta_s\tau_\vk}=
(i\gamma_{k}-\omega_{k})/h_{k}$. In deriving Eq. (\ref{initgrowth}), we also assumed $k_mR\gg\Delta_st$ and replaced
a slowly varying function $\tau_\vk\to\tau$.
   Expression (\ref{initgrowth}) describes
the initial formation of a wave packet \re{steadyOP1}.
Note that on a time scale $(q\Delta_s)^{-1}$ at which the order parameter deviation is of the order
$\sqrt{q}\Delta_s$, the width of the packet is   $l_p\approx\xi/\sqrt{q}$.

The observations above help to identify features of the post-threshold state
(\ref{steadyOP}) relevant for the experimental verification of our theory.
To be more specific, let us compute the correlator ${\cal K}({\vec r}_1-{\vec r}_2,t_1-t_2)=
\langle\delta\Delta({\vec r}_1,t_1)\delta\Delta({\vec r}_2,t_2)\rangle$, where
$t_1$ and $t_2$ are taken in the post-threshold stage. ${\cal K}({\vec r},t)$
characterizes the spatial and time distribution of the parametric \emph{noise} in the system.
Using the correlators for the random functions $e^{i\alpha_{\vk}}$ (see above), we obtain
from Eq. (\ref{steadyOP})
\beg\label{noise}
{\cal K}({\vec r},t)\propto {q\Delta_s^2}\frac{\sin(k_sr)\cos\Delta_s(t_1-t_2)}{k_sr},
\en
where $r=|{\vec r}_1-{\vec r}_2|$. This means that the spatial noise spectrum
$\propto\int{\cal K}({\vec k},\omega)d\omega$ has a
peak at the wave vector $k=k_s$ and decays as $1/k^2$ for $k\gg k_s$.

Formation of an isolated
wave packet (\ref{steadyOP1}) induces an oscillating supercurrent
${\vec j}_s\propto{\vec \nabla}\Phi({\vec r},t)$, where
$\Phi({\vec r},t)$ is the phase of the order parameter.
Setting $r_0=0$ we find that only the radial component of the current is nonzero.
To the lowest order in $q$,
${\vec j}_s({\vec r},t)\propto\hat{e}_r\sqrt{q}\cos(\Delta_st)[k_sr\cos(k_sr)-\sin(k_sr)]/(k_sr)^2$.
This implies a spatial re-distribution of Cooper pairs similar to the Friedel oscillations
in the density of a degenerate Fermi gas induced by a weak scattering potential.

In our discussion so far we  treated the pairing mode \re{dn}  giving rise
to the parametric instability as an independent external field. Inclusion of the feedback  on this mode
as weak turbulence ($q\ll 1$) develops   may modify the post-threshold state \re{steadyOP}.
We leave a detailed analysis of possible feedback effects for future studies.

In the post-threshold state (\ref{steadyOP}) the Fourier components of the order parameter
are $c_{\vk,\omega}\sim\delta(k-k_s)\delta(\omega-\Delta_s)$. Inelastic scattering or thermal effects generally leads to a broadening in
the momentum and frequency distributions of $c_{\vk,\omega}$ \cite{Lvov1976}.
The latter might cause a damping of the temporal oscillations in Eq. (\ref{steadyOP}).
On a time scale $t>\tau_\varepsilon$ dissipation
due to quasi-particle scattering processes ultimately
forces the system to reach an equilibrium state. Finally, we comment that in the
transient regime leading to an asymptotic state with constant $\Delta(t)=\Delta_s$
the order parameter
is $(\Delta(t)-\Delta_s)\propto\cos(2\Delta_st)/\sqrt{t}$ (cf. (\ref{dn})) \cite{classify}.
Oscillatory behavior suggests that this asymptotic state might never be attained
owing to the development of the parametric instability of the type considered above.

In conclusion, we have investigated the stability of the nonequilibrium asymptotic
states of a fermionic superfluid, which can be generated e.g. by a uniform
quench of the pairing strength.
We have demonstrated that in a system of size $L$ larger than the coherence length $\xi$ the asymptotic state (\ref{dn}) with periodic in time order parameter is unstable with respect to spatial fluctuations. The instability is due to the parametric excitation of two pairing modes with opposite momenta. The initial exponential growth of deviations from the homogeneous state
(\ref{initgrowth})
is suppressed by nonlinear effects eventually leading to a spatially nonuniform post-threshold state
described by Eq. (\ref{steadyOP}).
This state can be interpreted as a superposition of bubbles of the superfluid order
parameter (\ref{steadyOP1}) with random amplitudes.  The parametric instability of the uniform oscillations can be experimentally probed by spectroscopic noise measurements.

M. D.'s research was financially supported by the Department of Energy, grant
DE-FE02-00ER45790. E.A.Y. acknowledges the  financial support
by a David and Lucille Packard Foundation Fellowship for Science and Engineering, NSF
award NSF-DMR-0547769, and Alfred P. Sloan Research Fellowship. B.L.A. thanks
The US-Israel Binational Science Foundation for the financial support.


\begin{thebibliography}{99}

\bibitem{Ruutu1995} V. M. H. Ruutu \emph{et al.}, Nature (London) {\bf 382}, 334 (1995).

\bibitem{Warner2005} G. L. Warner and A. J. Leggett, Phys. Rev. B {\bf 71}, 134514 (2005).

\bibitem{oil2006} S. Casado, W. Gonz\'{a}lez-Vinas, and H. Mancini,
Phys. Rev. E {\bf 74}, 047101 (2006).

\bibitem{BEC2006} L. E. Sadler \emph{et al.}, Nature (London) {\bf 443}, 312 (2006).





\bibitem{Lvov1975} V. E. Zakharov, V. S. L'vov and S. S. Starobinets, Sov. Phys. -
Uspekhi {\bf 17}, 896 (1975).

\bibitem{LandafshitzI} L. D. Landau and E. M. Lifshitz, \emph{Classical Mechanics}, (Pergamon Press, London,1975).

\bibitem{Ruben1973} V. S. L'vov and A. M. Rubenchik, Sov. Phys. - JETP {\bf 37}, 263 (1973).

\bibitem{Lvov1976} V. S. L'vov, Sov. Phys. - JETP {\bf 42}, 1057 (1976).

\bibitem{Levitov2004} R. A. Barankov, L. S. Levitov and B. Z. Spivak,
Phys. Rev. Lett. {\bf 93}, 160401 (2004).


\bibitem{classify} E. A. Yuzbashyan, O. Tsyplyatyev and B. L. Altshuler, Phys. Rev. Lett. {\bf 96}, 097005 (2006).


\bibitem{Levitov2006} R. A. Barankov and L. S. Levitov, Phys. Rev. Lett. {\bf 96}, 230403 (2006).

\bibitem{Emil2006} E. A. Yuzbashyan and M. Dzero, Phys. Rev. Lett. {\bf 96}, 230404 (2006).

\bibitem{clean} In this paper we consider only the case of a clean superfluid.

\bibitem{Dzero2007} M. Dzero, E. A. Yuzbashyan, B. L. Altshuler and P. Coleman, Phys. Rev. Lett.
{\bf 99}, 160402 (2007).

\bibitem{Zakharov1971} V. E. Zakharov, V. S. L'vov and S. S. Starobinets, Sov. Phys. - JETP {\bf 32}, 656 (1971).

\end{thebibliography}
\end{document}